\documentclass[draft]{agujournal2019}
\usepackage{url} 
\usepackage{lineno}
\usepackage{graphicx}
\usepackage{rotating}
\usepackage{amsmath,amssymb}
\usepackage[inline]{trackchanges} 
\usepackage{soul}

\newif\ifnotesw \noteswtrue

\draftfalse

\journalname{}

\begin{document}

%
%


\title{Characterizing the impacts of multi-scale heterogeneity on solute transport in fracture networks}

%
%




\authors{Matthew R. Sweeney\affil{1}, Jeffrey D. Hyman\affil{1}, Daniel O'Malley\affil{1}, Javier E. Santos\affil{1,2}, J. William Carey\affil{1}, Philip H. Stauffer\affil{1}, Hari S. Viswanathan\affil{1}}

\affiliation{1}{Energy and Natural Resources Security Group (EES-16), Earth and Environmental Sciences Division, Los Alamos National Laboratory}
\affiliation{2}{Center for Nonlinear Studies, Theoretical Division, Los Alamos National Laboratory}




\correspondingauthor{Matthew R. Sweeney}{msweeney2796@lanl.gov}



\begin{keypoints}
\item We investigate the influence of fracture-to-fracture aperture heterogeneity in fracture networks on flow and transport using discrete fracture network modeling with constant pressure boundary conditions 
\item High aperture heterogeneity across a sparse fracture network results in counter-intuitive flow and transport behavior where flow rates decrease and solute transport time increases despite aperture increasing  
\item These effects of fracture-to-fracture aperture heterogeneity are limited at high network densities
\end{keypoints}

%
%

%
%


\begin{abstract}
We model flow and transport in three-dimensional fracture networks with varying degrees of fracture-to-fracture aperture/permeability heterogeneity and network density to show how changes in these properties can cause the emergence of anomalous flow and transport behavior. If fracture-to-fracture aperture heterogeneity is increased in sparse networks, velocity fluctuations can inhibit high flow rates and solute transport can be delayed, even in cases where hydraulic aperture is monotonically increased. As the density of the networks is increased, more connected pathways allow for particles to bypass these effects. We discover transition behavior where with relatively few connected pathways in a network from inflow to outflow boundaries, the first arrival times of particles are not heavily affected by fracture-to-fracture aperture heterogeneity, but the scaling behavior of the tails is strongly influenced due to the particles being forced to sample some of the heterogeneity in the velocity field caused by aperture differences. These results reinforce the importance of considering multi-scale effects in fractured systems and can inform flow and transport processes in both natural and engineered fracture systems, especially the latter where high aperture fractures are often stimulated and connect to existing fracture networks with smaller apertures.
\end{abstract}

\section*{Plain Language Summary}
Fractured rocks are important to study to understand subsurface geology, hydrology, and engineered systems. The amount of space in an individual fracture is closely correlated with how much fluid can flow within them, and subsequently, the transport capacity of the fracture. Individual fractures often form connected networks with other fractures, which complicates the prediction of flow and transport behavior because it is not easy to predict how fluids behave in such networks without demanding high-fidelity computational modeling. In this work, we study how changes in fracture network properties such as the variability of fracture apertures and the number of fractures in a network, affect flow and transport observables, such as the amount of actively flowing network structure, as well as the arrival times of solutes. Our results point to rich behavior where there is a close link between how much variability exists between individual fracture apertures, the network density, and the behavior of the solute transport. 
When the fracture networks have relatively few fractures, increases in both the magnitude of the fracture apertures and the variability between individual apertures can cause unusual changes in the flow and transport, in particular causing delays in solute transport times. 
However, these effects are gradually lost as the number of fractures in the network is increased.
Our results can inform both natural hydrological processes, as well as engineered systems, such as enhanced geothermal systems and hydraulic fracturing because these systems often create high aperture fractures that connect to natural fractures with lower aperture.

%
%

%


%
%
%
%

\section{Introduction}

Fractured rock formations are widely encountered in the subsurface, with fractures providing the main pathways for fluid flow and contaminant transport in low permeability rocks \cite{viswanathan2022fluid}. 
Understanding flow behavior in these systems is of great importance in various fields, including geology, hydrology, and environmental engineering, with specific applications ranging from CO$_2$ sequestration \cite{jenkins2015state} to the long term storage of spent nuclear fuel \cite{follin2014methodology,selroos2002comparison}. 
The hydraulic aperture of a fracture is a critical parameter controlling fluid flow in connected fracture networks because the individual apertures determine the hydraulic conductivity of each fracture, and therefore, influence the overall permeability of the connected network \cite{lang2014permeability,witherspoonwang,Witherspoon1980,zimmerman1996hydraulic}. 
Larger aperture fractures provide less resistance to flow, i.e., larger hydraulic conductivity, and conventional wisdom suggests that, larger apertures ought to therefore lead to higher volumetric flow rates, faster fluid flow and transport of contaminants. 
However, such changes in fluid flow properties require corresponding changes in the driving force of the system. 
Consider the case of a single fracture where inflow and outflow boundaries are prescribed constant flow rate boundaries (mass or volumetric).
Under these conditions, a uniform decrease in aperture is accompanied by a corresponding increase in the flow velocity due to the fact that the total amount of fluid flowing through the fracture is fixed by the boundary but the total volume decreases. 
Given less space, the fluid must increase velocity up due to volume conservation. 
Similarly, a uniform increase in the fracture aperture is accompanied by a corresponding decrease in the flow velocity for consistent reasons. 
Conversely, under Dirichlet, or constant pressure boundary conditions, each boundary of the fracture is held at a prescribed pressure for flow to be solved. 
According to Darcy's Law, an increase in the aperture must result in an increase in the volumetric flux and corresponding velocity because the pressure gradient does not change under these boundary conditions. 
What is unclear, however, is the interplay and feedback between changes in hydraulic aperture within the context of an interconnected network, and in particular when pressure is the primary driving force as is a common operational setup, which is what we consider and address here.

Previous research has shown that the relationship between individual fracture aperture, aperture distribution, and flow and transport behavior at the network scale is complex and depends on the specific properties of the network~\cite{national1996rock,national2020characterization}. 
Multi-scale heterogeneities are intrinsic aspects of fracture networks in nature and have been shown to play a key role in network scale flow and transport properties. 
Relevant length scales range from sub-fracture size, such as those related to surface roughness and asperities, to inter-fracture aperture heterogeneity, to network-scale attributes such as fracture topology and intensity.
Most studies have concluded that in-fracture aperture variability, whether natural or caused by external stress, can have significant effects on fluid flow and transport in a single fracture~\cite{blair1998,cardenas2007navier,DANG2019,hyman2021scale,kang2016emergence,WANG2016,zhou2018emergence,dreuzy2012influence,frampton2019advective,makedonska2016evaluating,PANDEY2015111}. 
Within the context of network flow, there is substantially less research than at the single fracture scale~\cite{hyman2016fracture,hyman2020characterizing,hyman2019linking,kang2020,kang2019,lei2017use,garipov2016discrete,maillot2016connectivity,mcclure2016fully,tran,sweeney2020stress}.
Nonetheless, a common finding amongst these studies is that when the aperture distribution widens, e.g., due to different fractures opening and closing from stress, flow channelization increases.
Yet, the theoretical understanding for these changes and their impact on upscaled transport properties has not been addressed.
The interdependence of multi-scale physical and hydraulic properties with anomalous flow and transport behavior has also been well documented by many previous researchers \cite<e.g.,>{berk97,edery2016structural,kang2020,kang2019stress,kang2017anomalous,Kang2015anomalous,roubinet2013particle,sherman2021review,wang2021channeling}.

In this paper, we investigate the effects of fracture-to-fracture aperture heterogeneity on fluid flow and transport in the context of complex fracture networks.
We begin with an analytic derivation providing a counter-intuitive prediction that the highest flow rates and fastest travel times in networks occur when inter-fracture aperture heterogeneity is low, not necessarily those networks with higher apertures as previously thought. 
We then demonstrate, explore, and characterize these effects at the network scale using a series of high-fidelity three-dimensional discrete fracture network (DFN) models coupled with  flow simulations and Lagrangian particle tracking.  
We observe that in sparse fracture networks, increasing the heterogeneity of hydraulic aperture distribution of fractures actually decreases the travel time of solutes through the network if the aperture changes are great enough. 
However, these effects are homogenized in higher density networks. 
This mitigation of meso-scale variation impacting macro-scale transport behavior is due to the availability of more percolation paths through the networks. 
In other words, the decreased constraint imposed on the flow by the network structure allows for a higher probability of connected fractures with similar apertures to exist, but there is transition behavior between these end-members.


\section{Theoretical Background}

We begin by considering a series of three fractures where flow is driven by Dirichlet pressure boundary conditions and steady-state flow and solute transport times can be solved analytically for this system using Darcy's Law.
We choose this problem because it isolates how changing a single fracture aperture in a series affects flow and transport at the network scale. 
Consider first the cross sectional flow rate in the network, neglecting gravity, given by
\begin{linenomath*}
\begin{equation}
    Q=\frac{b_h^3}{12\nu}\frac{p_0-p_1}{\rho L}
\end{equation}
\end{linenomath*}
where $\nu$ is kinematic fluid viscosity, $p_0$ and $p_1$ are boundary pressures, $\rho$ is fluid density,
 \begin{linenomath*}
\begin{equation}
    L=l_1+l_2+l_3
\end{equation}
\end{linenomath*}
and the aperture of the network is given by the harmonic average of the individual apertures, which are denoted by the subscript $i=1,2,3$,
 \begin{linenomath*}
\begin{equation}
    b_h^3=\frac{3}{\frac{1}{b_1^3}+\frac{1}{b_2^3}+\frac{1}{b_3^3}}.
\end{equation}
\end{linenomath*}
We can then calculate the velocities in each of the fractures
 \begin{linenomath*}
\begin{equation}
    v_i=\frac{Q}{b_i}
\end{equation}
\end{linenomath*}
and the advective travel time in each fracture is
 \begin{linenomath*}
\begin{equation}
    t_{f_i}=\frac{l_i}{v_i}
\end{equation}
\end{linenomath*}
Thus the total advective travel time is
 \begin{linenomath*}
\begin{equation}
    t_f=t_{f_1}+t_{f_2}+t_{f_3}.
\end{equation}
\end{linenomath*}

If we choose values for the unknowns in this system, namely, $l_i, p_0, p_1, \rho$ and $\nu$, we can explore the behavior further. 
Assume that $l_1=l_2=l_3=1$ m, $p_0=1.1$ MPa, $p_1=1.0$ MPa, $\rho=997$ kg/m$^3$, and $\nu=8.93\times10^{-7}$ m$^2$/s.
Without loss of generality, assume the first and third fractures in the series have apertures of $10^{-4}$ m and we systematically vary the aperture in the second fracture from $2\times10^{-5}$ m to $5\times10^{-3}$ m. For each aperture, we solve the analytical solution to determine the advective travel times of particles through the network. 

\begin{figure}
    \begin{center}
    \noindent\includegraphics[width=0.75\textwidth]{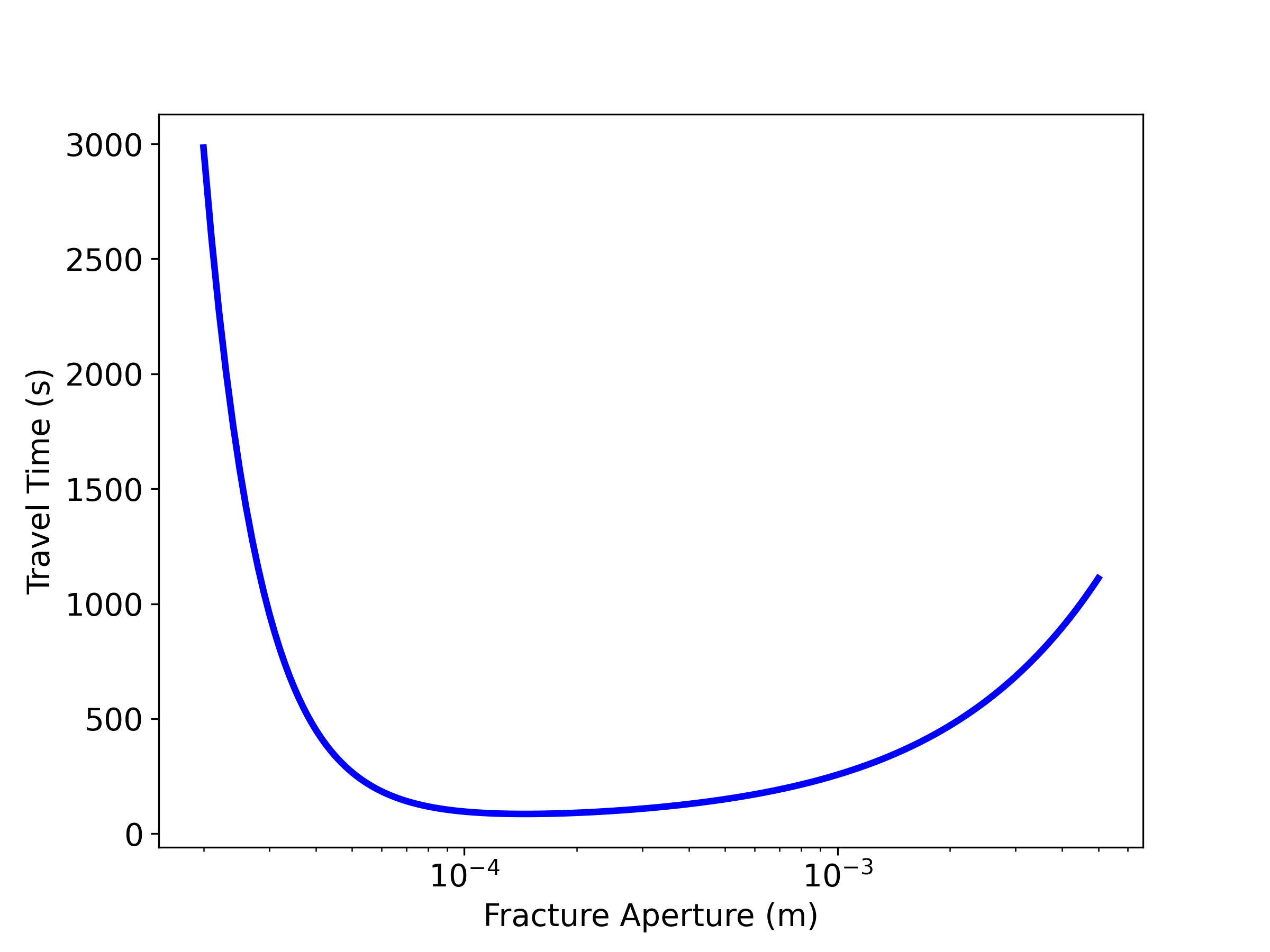}
    \caption{Analytical solution for advective particle travel time as a function of the second fracture aperture in a three fracture series. The first and third fractures in the series have the same aperture of $10^{-4}$ m.}
    \label{3frac_line}
    \end{center}
\end{figure}

Figure \ref{3frac_line} shows the results plotted as a function of the aperture of the second fracture. 
Any  decrease \textit{or} increase beyond a negligible amount in the aperture of the middle fracture results in a delay in the particle breakthrough times relative to the case where the middle fracture has the same aperture as the left and right fractures. So even if the system has a larger harmonic average due to that larger second fracture apertures relative to $10^{-4}$ m, there is a delay due to the velocity decreasing in the second fracture caused by the larger aperture of the second fracture. 
One can make a similar argument using Bernoulli's principle in pipe flow where a large pipe is connected to a smaller pipe, driven by an excess pressure \cite{levi}. Attempting to increase the flow rate out of the second pipe by changing the size of the first is a futile task as the velocity limit of the system occurs when the pipes are equal diameter. The question then becomes, how does such a phenomenon manifest at network scale in fracture networks and is it possible that networks with greater permeability have slower transport times through them? 

\section{DFN Simulations}
We use the {\sc dfnWorks}~\cite{hyman2015dfnworks} software suite to generate, mesh, and resolve flow and particle transport on three-dimensional fracture networks with increasing complexity. 
Comprehensive details and methods of the discrete fracture methodology and {\sc dfnWorks} are in the Supplementary Information.
The networks we consider are "semi-generic" meaning that they do not represent a particular field site, but their attributes are loosely based on field observations. 
Once the networks are generated and meshed, we solve for the steady-state pressure field and volumetric flow rates of an incompressible fluid by invoking Dirichlet boundary conditions using {\sc pflotran} \cite{lichtner2015pflotran}, which uses a two-point flux finite volume approach. 
The governing equation we adopt is Reynold's equation, which is linear in the pressure gradient. 
Solute transport within each DFN is simulated using a Lagrangian approach, where a plume of purely advective, non-reactive particles trace pathlines through the fluid velocity field within the network \cite{makedonska2015particle}. 
Each simulation in the following sections uses 100,000 particles and their initial locations are uniformly placed along the fractures that intersect the inflow boundary plane.
The travel times of particles exiting the domain is recorded and the probability density function of the travel time distribution (i.e., the breakthrough curve) is empirically constructed. 
Further details of the flow and transport approach, including governing equations, can also be found in the Supplementary Information.
In order that we retain a singular focus to this study, we assume that the apertures follow a smooth parallel plate assumption.
Recent advances allow for in-fracture aperture variability to be incoroporated into DFN network scale models, but our goal is to isolate changes at the fracture scale not smaller scales, so we assume that fracture apertures are uniform within individual fractures.


\section{Results}

\subsection{Single Percolation Path}
\begin{figure}
\noindent\includegraphics[width=\textwidth]{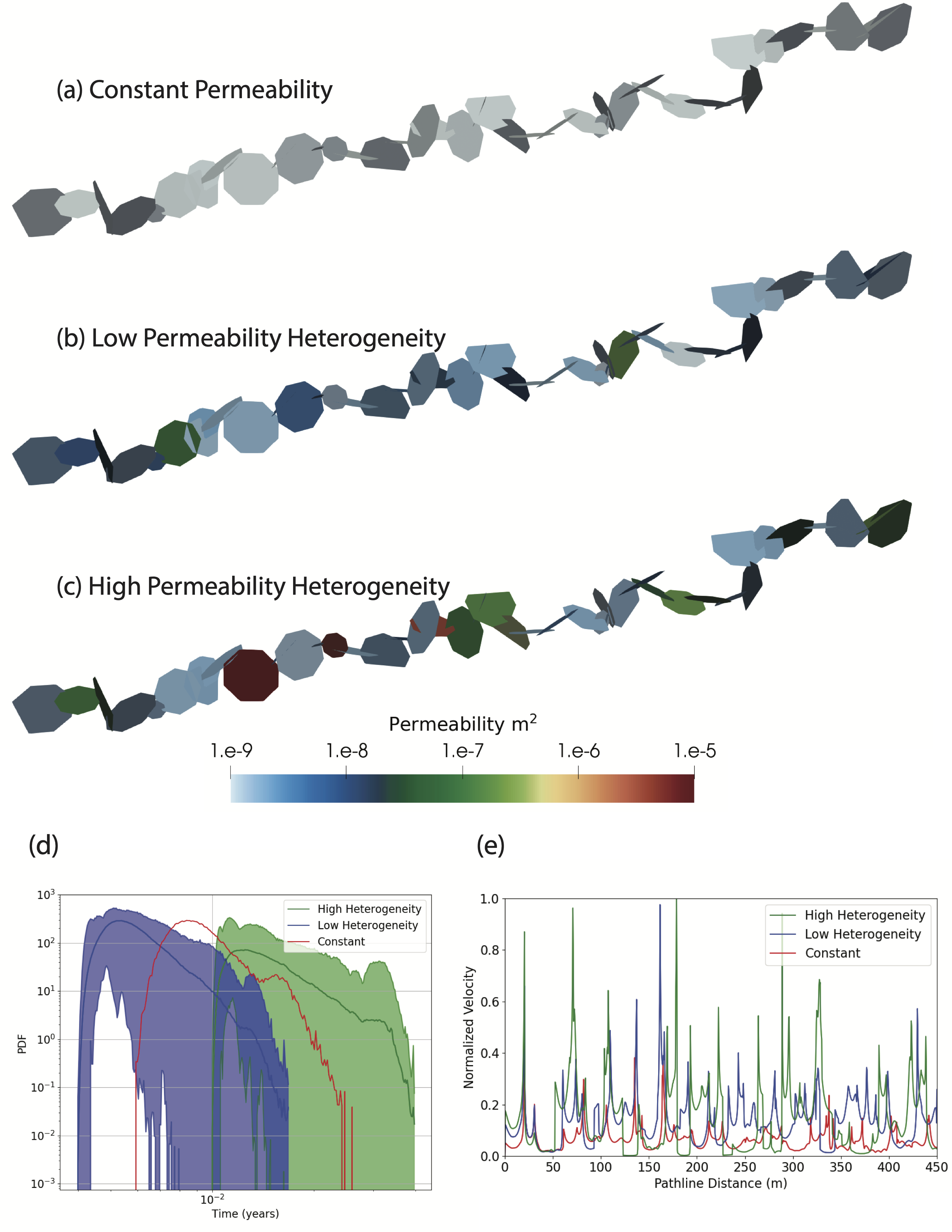}
\caption{(a) DFN with constant fracture apertures; (b) DFN with relatively low fracture-to-fracture aperture heterogeneity; (c) DFN with relatively high fracture-to-fracture aperture heterogeneity; (d) Particle breakthrough curves for ensembles of DFNs shown in (b) and (c), as well as constant case (a); (e) Velocity as a function of pathline distance for randomly selected particles in each type of DFN.}
\label{43frac}
\end{figure}

In this section, we consider DFNs with the same 43 fractures and geometry, with only one percolation path through the network (Figure \ref{43frac}). By doing this, we can isolate the effects of changing and increasing apertures on flow and transport along particle trajectories forced to enter and exit each fracture in the DFN and assess if and how the phenomenon described in Section 2 translates to systems with many more fractures.

The DFN is generated using a truncated power law distribution with radii $r$ and a decay exponent $\alpha$, which is a commonly measured field quantity \cite{bonnet2001scaling},
\begin{linenomath*}
\begin{equation}
    p_r=(r,r_0,r_u)=\frac{\alpha}{r_0}\frac{(r/r_0)^{-1-\alpha}}{1-(r_u/r_0)^{-\alpha}}
\end{equation}
\end{linenomath*}
where $r_u$ and $r_0$ are the upper and lower radius cutoffs, respectively. We use values $r_u=10.0$ m, $r_0=2.0$ m, and $\alpha=2.0$. The fractures are represented by randomly oriented discs, mimicking a disordered fractured media~\cite{hyman2017dispersion}.
The same flow problem is solved as the following sections in this paper. Namely, Dirichlet pressure boundary conditions, with an inflow pressure of 2.0 MPa and outflow pressure of 1.0 MPa. 
Because the governing equations are linear, modifying the pressure values does not change the phenomenology of the flow and transport in terms of the velocity field structure, but merely shifts particle arrival times in one direction or the other depending on how the gradient is changed.

Figures \ref{43frac}a,b,c show three DFNs under consideration in this section. The total domain length is 300 m and each non-boundary fracture is intersected by exactly two other fractures, while the boundary fractures are connected by one other fracture and the adjacent inflow or outflow boundary. 
Three different suites of fracture apertures are considered. 
First, those where all fractures have the same hydraulic aperture $b_0 = 1.1\times10^{-4}$. 
Then, two suites of realizations where we only allow fracture apertures greater than $b_0$.
We restrict the aperture to be higher because we are mostly interested in the phenomenology described in Section 2. 
We accomplish this by initially generating an aperture distirbution using a log-normal distribution, with defined mean and variance. 
If an aperture value that is generated results in a aperture less than $b_0$. we simply flip the sign of the difference and add it to $b_0$, hence we end up with apertures greater than $b_0$.
Since the mean fracture aperture varies each realization as a result of this generation approach, we generate 30 networks for each suite. 
We quantify this heterogeneity using the coefficient of variation, $CV=\sigma/\mu$, where $\sigma$ is the population standard deviation and $\mu$ is the population mean. 
One ensemble of 30 realizations has $CV=17.73$, and the other ensemble has $CV=6.17$, which we hereafter refer to as \textit{relatively} high and low heterogeneity, respectively, noting these values are relative to each other and not a particular field site.

Figure \ref{43frac}d shows the BTCs for the high heterogeneity, low heterogeneity, and constant aperture transport simulations, shaded with 95\% confidence intervals for the former two. When the aperture heterogeneity between the individual fractures is low, there is early first arrival of particles relative to the constant aperture case. However, when the heterogeneity is increased, there is a delay in arrival relative to the constant case, even though in both suites there are no fractures with apertures less than $b_0$ meaning a more permeable network exhibits higher solute residence times than a less permeable network due to differences in fracture-to-fracture aperture.

Figure \ref{43frac}e shows normalized particle velocities plotted as a function of pathline distance for three individual particle trajectories (one randomly selected from each network suite). Note the trajectories are not necessarily one-to-one comparisons because particles are not guaranteed to take the same trajectory through an individual fracture or experience the same fractures at the same pathline distance. Several striking features are evident. 
Each trajectory shows considerable velocity variability corresponding to changes in fracture residency. 
The constant aperture network shows the least amount of variability relative to the other cases. The particle trajectory from the high heterogeneity network records the highest velocities, but also the lowest, reaching values close to zero several times. 
The particle trajectory from the low heterogeneity network is consistently higher than the constant case, and does not experience the same dramatic drops close to zero velocity that the higher heterogeneity case does.

These results extend upon the the phenomenology discussed in Section 2 with the three fracture series and elucidate the expected variability depending on the actual fracture aperture encountered in different networks. 
In summary, if particles are confined to a single percolation path in a sparsely fractured network, the transport times are closely linked to the variation in fracture-to-fracture aperture. 

\subsection{Multiple Percolation Paths}
In this section, we consider DFNs with multiple percolation paths, but with different densities.
First, we consider a DFN with relatively low density/few percolation paths, which we define here as any connected pathway of fractures that connects the inflow and outflow boundaries, and then we consider a DFN with relatively high density/many percolation paths. 
Note that "high" and "low" are used as relative terms for the networks considered and are not meant to be quantiative outside of the comparison presented herein.

To generate these networks, we again employ a truncated power law distribution governed by eq. (7), where the fractures are represented as planar ellipses. For these networks we assume $\alpha=1.8$, $r_0=1$ m and, and $r_u=10$ m. The networks are generated in a cubic domain with sides of length 50 m. For consistency, we again refer to a constant baseline aperture case where all of the fractures have constant aperture of $b_0 = 1.1 \times 10^{-4}$, but also generate ensembles of 10 realizations each with relatively high and low aperture heterogeneity, but again restrict the fracture apertures to only be greater than $b_0$ using the same approach described in Section 4.1.

Similar to previous work \cite<e.g.,>{hyman2020characterizing}, we measure the density of the networks using the percolation parameter $p$ as defined by \citeA{dreuzy2012influence}
\begin{linenomath*}
\begin{equation}
    p=\frac{N}{L^2}\int_{r_0}^{r_u}dr\textrm{min}(r,\alpha L)p_r(r).
\end{equation}
\end{linenomath*}

For this set of generation parameters, the critical percolation threshold, $p_c$, defined as the minimum number of fractures required such that there is a connected cluster of fractures that spans the entire domain, is $\approx 750$ fractures. We can further define a non-dimensional percolation parameter, $p'=p/p_c$ where $p$ is the number of fractures inserted into the computational domain during network generation. We only consider DFNs where the entire network connects the inflow and outflow boundaries, i.e., isolated fractures and clusters of fractures are not retained in the final networks. Hence, the final number of fractures in the DFNs is not equivalent to $p$. Nevertheless, $p'$ is a useful metric because it places all the networks into a relative context \cite{hyman2020characterizing}.

\subsubsection{Low Density, $\boldsymbol{p'=3}$}
\begin{figure}
\noindent\includegraphics[width=\textwidth]{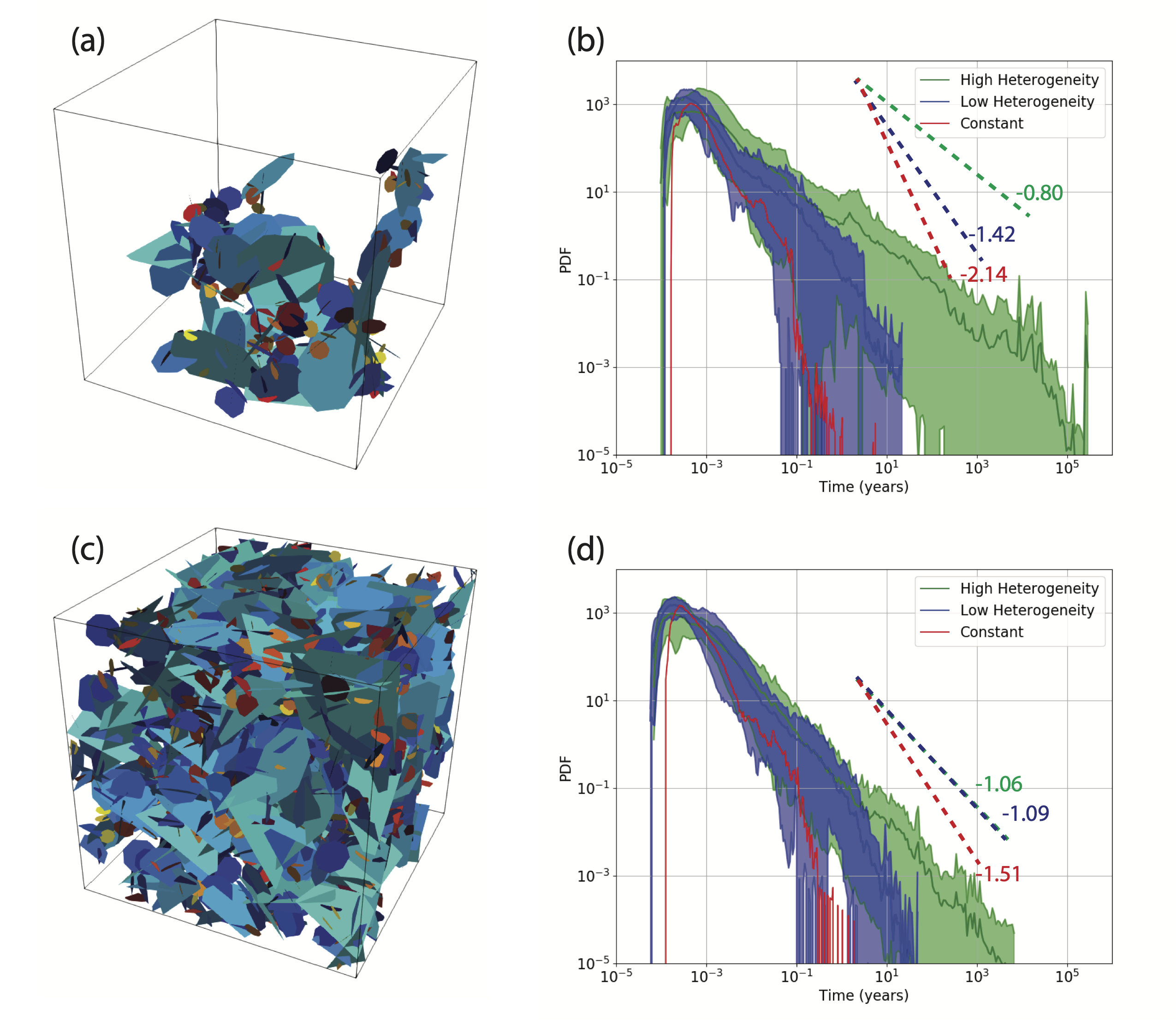}
\caption{(a) $p'=3$ DFN (b) Particle breakthrough curves for ensembles of high and low aperture heterogeneity DFN shown in (a). 10 realizations of each high and low were generated and the curves are shaded with 95\% confidence intervals. (c) $p'=5$ DFN (d) Particle breakthrough curves for ensembles of high and low aperture heterogeneity DFN shown in (c). 10 realizations of each high and low were generated and the curves are shaded with 95\% confidence intervals.}
\label{multi}
\end{figure}

For the relatively low density DFNs, we choose a value of $p'=3$, which results in an initial seed of 2250 fractures in the generation. Using a randomly initialized seed, we end up with a DFN with 184 fractures after isolated fractures are removed Figure \ref{multi}a. Similar to Section 4.1, we generate two ensembles of the same network structure with different fracture-to-fracture aperture heterogeneities where the apertures are only allowed to be greater than $b_0$. For these networks, we generate 10 realizations of each. 
The relatively high heterogeneity ensemble has $CV=35.07$ and the relatively low heterogeneity ensemble has $CV=9.74$. We also include an example with constant aperture and the same flow and transport problem is solved as in Section 4.1.

Figure \ref{multi}a shows the network under consideration colored by fracture number.
It is clear that there are more percolation paths through the network compared to the network in Section 4.1., but still relatively few compared to the network in the next section. Figure \ref{multi}b shows the BTCs for the high heterogeneity, low heterogeneity, and constant DFN realizations. Both high and low ensembles show similar first arrival times, which are nearly the same as the constant aperture network. 
The scaling of the tail in the BTCs are quite different, however (Figure \ref{multi}b). The high heterogeneity DFNs show heavy/long tails that decay quite slowly (slope -0.80) relative to both the low heterogeneity DFNs (slope -1.42) and the constant aperture DFN (slope -2.14). The tails of the low heterogeneity DFN BTCs show scaling behavior in between and different than the constant DFN and high heterogeneity DFNs. 

From an Eulerian point of view we can investigate how much of the domain is actively flowing using the flow channeling density indicator $d_Q$ presented by~\cite{maillot2016connectivity}
 \begin{linenomath*}
 \begin{equation}\label{eq:flow_channeling}
d_Q = \frac{1}{V} \cdot \frac{ (\sum_f \cdot S_f \cdot Q_f)^2}{(\sum_f \cdot S_f \cdot Q_f^2)}~\,
\end{equation}
 \end{linenomath*}
whose definition is inspired by the participation ratio developed in solid state physics \cite{bell1970localization,edwards1972numerical} and has been adapted for use in the geosciences as well~\cite{andrade1999inertial,davy1995localization,davy2010likely}.
In eq. \eqref{eq:flow_channeling},  $Q_f$ is one-half the absolute value of the total volume of fluid exchanged by a fracture $f$ with its neighbors and $S_f$ is the fracture surface area and $V$ is the total size of the domain. 
This measure is very similar to the fracture intensity (total fracture surface area per unit volume), defined by 
 \begin{linenomath*}
 \begin{equation}\label{eq:p32}
P_{32}= \frac{\sum_f \cdot S_f}{V} ~,
\end{equation}
 \end{linenomath*}
Note that if $Q_f$ is uniform across all fractures, then $d_Q = P_{32}$. 
Comparing  eq. \eqref{eq:flow_channeling}  with eq. \eqref{eq:p32}  shows that $d_Q$ is a measure of \emph{active $P_{32}$} or \emph{flowing $P_{32}$} and the ratio $d_Q/P_{32}$ is the percentage of the network with significant flow.

For the low density DFNs, this value does not change greatly, with mean $d_Q/P_{32}$ values of 0.352, 0.360, and 0.375 for the high heterogeneity, low heterogeneity, and constant DFNs, respectively. The logical extension of this result is that the particles in the different DFN suites are experiencing the same fractures through the low density network as they are transported.

\subsubsection{High Density, $\boldsymbol{p'=5}$}
For the relatively high density DFNs, we choose a value of $p'=5$, which results in an initial seed of 3750 fractures in the generation. Using a randomly initialized seed, we end up with a DFN with 1428 fractures after isolated fractures are removed (Figure \ref{multi}c). 10 realizations of relatively high and low inter-fracture aperture heterogeneity networks are again generated, along with a network with constant fracture aperture. For these networks, the high heterogeneity ensemble has $CV=31.78$ and the relatively low heterogeneity ensemble has $CV=5.97$. 

Figure \ref{multi}d shows the BTCs for the high heterogeneity, low heterogeneity, and constant DFN realizations for the relatively dense network. Both high and low ensembles again show similar first arrival times, which are slightly earlier than the constant aperture network, but not substantially so. The scaling of the tail in the BTCs are now quite similar between the high and low heterogeneity DFNs, with slopes of -1.06 and -1.09, respectively. The BTC of the constant DFN decays faster than both the high and low heterogeneity DFNs, but is within the 95\% confidence intervals of both up until $\approx 10^{-1}$ years. For the high density DFNs, mean $d_Q/P_{32}$ values decrease as $CV$ is increased, indicating an increase in flow channeling. The $d_Q/P_{32}$ values for the high heterogeneity, low heterogeneity, and constant DFNs are 0.225, 0.278, and 0.326, respectively.

\section{Discussion and Conclusions}
We have investigated the impact of fracture-to-fracture aperture heterogeneity on flow and transport in variable density fracture networks using high fidelity flow modeling and Lagrangian particle tracking in DFN models. Our results point to several interesting points previously not discussed in the fracture literature or that might be re-interpreted in light of the findings of this study.

In terms of transport, when particles are forced to sample the extent of the fracture-to-fracture aperture heterogeneity, there is a significant effect on both the timing of the initial breakthrough and the tailing behavior of the particle breakthrough that could be classified as anomalous. As the fracture network density, and subsequently the number of percolation paths, is increased, the initial breakthrough behavior is less dependent on the inter fracture aperture heterogeneity. Our results suggest that in the denser networks there is a greater degree of flow channeling as the $CV$ is increased, hence changes in the fracture-to-fracture aperture heterogeneity are accommodated by changes in flow channeling, which explains why the breakthrough behavior of particles is similar between high and low heterogeneity ensembles. When there are relatively few percolation paths, however, the tailing behavior shows a strong dependence on the inter fracture aperture heterogeneity, with heavy and long tails in cases with high $CV$. Increasing the fracture network intensity to the point where there are many percolation paths washes out most of the effects of inter-fracture heterogeneity because the particles are no longer forced to sample the aperture heterogeneity present in the networks. At the extreme case where there is only one percolation path, the effects of aperture heterogeneity are amplified to the point where initial breakthrough times are heavily dependent on the degree of variation, here measured by $CV$. When apertures are increased in networks with a single percolation path, it is possible to actually delay the solute first arrival times due to strong velocity fluctuations from high fracture-to-fracture aperture changes, which is consistent with counter-intuitive analytical expectations. We only considered networks where the fracture apertures were increased relative to a constant baseline to probe this behavior.

Previous research has studied the link between fracture-to-fracture aperture heterogeneity from different perspectives than what we have done here and could be re-interpreted in light of our results. For one aspect, it has been published that under anisotropic mechanical stress fields, aperture distributions may widen and become more variable \cite{kang2019,lei2017use,sweeney2020stress,garipov2016discrete,mcclure2016fully,tran}. If the network is short circuited via connected shear fractures, one can expect early arrival times of solutes relative to the original network. We suggest this is less a consequence of apertures actually increasing, but the fact that given equivalent stress conditions, the connected fractures are likely similar in apertures, which would result in a low aperture heterogeneity pathway through the network, especially in sparser networks with few percolation paths. A second aspect to consider is with respect to flow channeling. Separate studies have looked at the relationship between flow channeling and network density \cite{hyman2020characterizing} and fracture-to-fracture aperture heterogeneity \cite{hyman2016fracture,kang2020}. In the case of network density, the relationship between flow channeling and network density was shown to be negatively correlated, while for fracture-to-fracture aperture heterogeneity the relationship is positively correlated. Our results show that at low network density, the degree of flow channeling is relatively unaffected by changes in the fracture-to-fracture aperture heterogeneity, whereas at high network density flow channeling does indeed increase, which is consistent with \citeA{kang2020}. As alluded to in the previous paragraph, this is an important aspect to consider because the degree of flow channeling tells us how much of the network velocity field the particles are sampling as they traverse the network structure. In the case where the network density is high and the aperture heterogeneity is also high, the increase in flow channeling signifies that much of the network is not actually contributing to the flow and transport. Consequently, changes in the fracture apertures are not likely to have as dramatic effects as they do in sparser networks.

Changes in both fracture-to-fracture aperture heterogeneity and network density have strong effects on flow and transport in fracture networks, which we characterized using a flow channeling metric and breakthrough behavior of particles. This study was limited to theoretical considerations, but could be extended to natural or engineered fractured systems in the future, which could elucidate these relationships further. Much of the energy production around the world is from subsurface fractured systems \cite{viswanathan2022fluid}. Tapping into these energy sources often involves stimulating them with high aperture fractures, which then connect to natural fracture networks. Our results suggest this should be done carefully because creating a system with large differences in fracture aperture could have the opposite of the intended effect in terms of energy production.

\section*{Open Research Section}
The software used for the simulations, dfnWorks, is open source and can be obtained at http://dfnworks.lanl.gov. Run scripts to generate reproduce data from these simulations are in the Mendeley repository http://url. Los Alamos National Laboratory unclassified release number: LA-UR-23-25773.

\acknowledgments
M.R.S., J.D.H., D.M., J.W.C., and H.S.V. would like to acknowledge support from the Department of Energy Basic Energy Sciences program (LANLE3W1) for support. M.R.S., J.D.H., and H.S.V. would like to acknowledge support from Los Alamos National Laboratory LDRD Award 20220019DR. M.R.S. would also like to thank support from Los Alamos National Laboratory LDRD Award 20220175ER. J.E.S. would like to thank the Center for Nonlinear Studies for support. Los Alamos National Laboratory is operated by Triad National Security, LLC, for the National Nuclear Security Administration of U.S. Department of Energy (Contract No. 89233218CNA000001).

%
%


%
%
%
%
%

\bibliography{paper.bib}
\end{document}


%
%


\title{Supporting Information for "Characterizing the impacts of multi-scale heterogeneity on solute transport in fracture networks"}
%
%

%
%



\authors{Matthew R. Sweeney\affil{1}, Jeffrey D. Hyman\affil{1}, Daniel O'Malley\affil{1}, Javier E. Santos\affil{1,2}, J. William Carey\affil{1}, Philip H. Stauffer\affil{1}, Hari S. Viswanathan\affil{1}}


\affiliation{1}{Energy and Natural Resources Security Group (EES-16), Earth and Environmental Sciences Division, Los Alamos National Laboratory}
\affiliation{2}{Center for Nonlinear Studies, Theoretical Division, Los Alamos National Laboratory}


%
%

%

\begin{article}

%
%

\section{Flow and transport equations}
We use {\sc dfnWorks}~\cite{hyman2015dfnworks} to generate 3D-DFNs in this study. 
In the DFN approach, fractures are represented as 2D planar objects in 3D space, e.g., rectangles or discs, and those fractures interconnect to form a network through which flow and transport is modeled. Details of each DFN or realization of DFNs are provided in the main paper, but some of the general aspects of the approach are highlighted here. First, none of the networks in this study are meant to represent any particular site or rock, though the generation parameters are loosely based on those commonly measured in the field, so can be considered "semi-generic" in this regard. Each network is meshed using an unstructured conforming Delaunay triangulation using the feature rejection algorithm for meshing \cite<FRAM;>{hyman2014conforming}. Details of FRAM can be found in the original paper, as well as subsequent papers, so will not be recounted here. In each DFN, fractures are represented by planar ellipses whose radii $r$ follow a power law distribution with a decay exponent $\alpha$, which is a commonly measured field quantity \cite{bonnet2001scaling},
\begin{equation}
    p_r=(r,r_0,r_u)=\frac{\alpha}{r_0}\frac{(r/r_0)^{-1-\alpha}}{1-(r_u/r_0)^{-\alpha}}
\end{equation}
where $r_u$ and $r_0$ are the upper and lower radius cutoffs, respectively. We provide these values when the different networks are introduced in the results.

We model steady-state flow within DFNs using the Reynolds equation, which is valid due to the large contrast between fracture lengths and their hydraulic apertures~\cite{zimmerman1996hydraulic}
  \begin{linenomath*}
\begin{equation}\label{eq:2Dflow}
\nabla \cdot ( b^3 \nabla P ) = 0\,.
\end{equation}
 \end{linenomath*}
Countless studies have used this approach to calculate the pressure distribution in DFNs and it has been extensively validated.

Dirichlet boundary conditions are used in each simulation in this study and are applied such that a pressure difference is prescribed across the domain and aligned with the $x$ axis. 
We discuss the implications of these boundary conditions further in Section 3.
We do not consider the feedback between fluid pressure and any mechanical stress field. 
Neumann, no-flow, boundary conditions are applied along lateral boundaries and gravity is not included in any of the simulations.
The matrix surrounding the fractures is assumed to be impervious and there is no interaction between flow within the fractures and the solid matrix, i.e., matrix diffusion is not considered in these simulations. 
Under this simulation setup, the structure of the flow field does not change with changes in the applied pressure difference and the value used is therefore arbitrary.
The pressure distributions and volumetric flow rates are determined by numerically integrating eq. (\ref{eq:2Dflow}) using an unstructured two-point flux finite volume scheme implemented in {\sc pflotran} \cite{lichtner2015pflotran} The Eulerian velocity field within the DFN is then reconstructed from the volumetric fluxes and pressures using the method provided in~\citeA{makedonska2015particle} and \citeA{painter2012pathline}.

Transport through the DFNs is simulated using a plume of nonreactive indivisible particles that trace pathlines through the velocity field. 
Particles are injected over a plane $\Omega_a$ located at the inlet plane  perpendicular to the mean flow direction, where the  initial particle positions are denoted as $\va$.
The distribution of particles along the inlet plane is assigned using equal distance weighting such that each initial particle position is equally spread out along the inflow boundary fracture~\cite{hyman2015influence,kang2017anomalous,kreft1978on}. 
100,000 particles are used in each of the transport simulations except for the three fracture series in Section 3, which does not require that many particles due to the simplicity of the system.

The trajectory $\vx(t;\va)$ of a particle starting at ${\bf a}$ at time $t=0$ is given by the advection equation
 \begin{linenomath*}
\begin{align}\label{eq:trajectory}
\frac{d \vx(t;\mathbf a)}{d t} = \vv(t;\va), &&  \vx(0;\va) = \va,
\end{align}
 \end{linenomath*}
where the Langrangian velocity $\vv_t(t;\mathbf a)$ is given in terms of
the Eulerian velocity $\vu(\vx)$ as 
\begin{linenomath*}
    \begin{equation}
        \vv(t;\va) = \vu[\vx(t;\va)]. 
    \end{equation}
\end{linenomath*}
When particles enter fracture intersections, the dynamics of which are a subgrid scale process, we assume a complete mixing model where the probability to exit into a fracture is proportional to the outgoing volumetric flow rate~\cite{Berkowitz1994mass,stockman1997lattice,Park2001intersections,park2003transport,kang2015anomalous,sherman2018characterizing}. 

The length of the path line $\ell$ is used to parameterize the particle trajectory. 
The total travel time for a particle to pass through the domain $x \geq x_L$ is given by 
\begin{linenomath*}
    \begin{align}
        \label{taux}
        \tau(x_L;\va) = \int_0^{\ell(x \geq x_L)} \frac{d \ell}{\| v(\ell) \|}~. 
    \end{align}
\end{linenomath*}
The values of $\tau(x_L,\va)$ across the particle plume are combined to compute the relative solute mass flux $F(t,x_L)$ breakthrough at a time $t$, 
 \begin{linenomath*}
\begin{equation}\label{eq:CD}
F(t,x_L) = \int\limits_{\Omega_a} d \va~\delta[\tau(x_L,\va) - t],
\end{equation}
 \end{linenomath*}
where, $\delta(t)$ is the Dirac delta function. 
Equation~\eqref{eq:CD} is known as the particle breakthrough curve (BTC). 

\bibliography{bib.bib}

%
%
%
%
%

%
%
\end{article}
\clearpage